\documentclass[linenumbers]{aa}  

    \usepackage{graphicx}
    \usepackage{txfonts}
    \usepackage[colorlinks = true, allcolors=blue]{hyperref}
    \usepackage{stfloats}
    \usepackage{booktabs}
    \usepackage{orcidlink}
    \usepackage{sidecap}
    \usepackage{pdflscape}
    \usepackage[normalem]{ulem}
    \newcommand{\Msun}{\hbox{M$_\odot$}}

\begin{document}

       \title{A rotation-based census of blue lurker candidates in open clusters}
       % \subtitle{}
       \titlerunning{Rotation-based census of blue lurkers}
       % \authorrunning{Jadhav}
       
       \author{
            Vikrant V. Jadhav\orcidlink{0000-0002-8672-3300}\inst{1}, 
            Khushboo K. Rao\orcidlink{0000-0001-7470-9192}\inst{2},
            Elisabetta Reggiani\orcidlink{0009-0000-4254-7475}\inst{3,4}
        }
       \institute{
            Astronomical Institute, Faculty of Mathematics and Physics, Charles University, V Holešovičkách 2, CZ-180 00 Praha 8, Czech Republic\\
            \email{vikrant-vinayak.jadhav@matfyz.cuni.cz}
            \and
            Institute of Astronomy, National Central University, 300 Zhongda Road, Zhongli 32001 Taoyuan, Taiwan\\
            \email{khushboo@gm.astro.ncu.edu.tw}
            \and
            Dipartimento di Fisica e Astronomia, Università di Firenze,  Via G. Sansone 1, 50019 Sesto Fiorentino, FI, Italy\\
            \email{elisabetta.reggiani@unifi.it}
            \and
            INAF – Osservatorio Astrofisico di Arcetri, Largo E. Fermi 5, 50125 Firenze, Italy
            }
       \date{Received November 20, 2025; accepted January 28, 2026}
    
    % \abstract{}{}{}{}{} 
    % 5 {} token are mandatory
     
      \abstract
      % context heading (optional)
       {}
      % aims heading (mandatory)
       {
       Blue lurkers (BLs) are rejuvenated main-sequence stars hidden among normal main-sequence stars on color-magnitude diagrams of star clusters. In comparison, the blue straggler stars, formed via similar mass transfers or mergers, occupy a distinct space in the color-magnitude diagrams.
       We compile a list of BL candidates in open clusters using available rotation catalogs.
       }
      % methods heading (mandatory)
       {
       BLs can be identified using either unusually faster rotation compared to similar mass stars, which is a signature of recent accretion, or the presence of a companion, which can only be formed by mass donation, e.g., an extremely low mass white dwarf. Here, we searched for fast-rotating stars on the main sequence of open clusters using Kepler, TESS, and spectroscopic rotation indicators, such as rotation periods and $v\sin i$ measurements.
       }
      % results heading (mandatory)
       {
       We identified 97 new BL candidates across 35 open clusters, almost tripling the previously known sample of 36. Based on the estimated completeness of $\approx$3\%, thousands of BLs are likely hidden within the cluster population. Detailed spectroscopic and time-series analyses will be essential to confirm their mass-transfer histories.
       }
      % conclusions heading (optional)
       {}
    
    
       \maketitle
    %
    %-------------------------------------------------------------------

\section{Introduction} 

Stars that undergo mass exchange or mergers can experience rejuvenation, appearing younger and more massive than predicted by single stellar evolution. The most prominent examples of such systems are blue straggler stars (BSSs), which lie above and to the bluer side of the main-sequence turnoff (MSTO) in cluster color–magnitude diagrams (CMDs). They are understood to form through mass transfer in binaries \citep{McCrea1964MNRAS.128..147M} or stellar mergers or collisions \citep{Hills1976ApL....17...87H}.

A related but less conspicuous class, known as blue lurkers (BLs), represents main-sequence stars that have also undergone mass accretion or mergers but remain photometrically indistinguishable from normal main-sequence stars. BLs are therefore considered the main-sequence counterparts of BSSs (see Section~\ref{sec:definition} for a formal definition).

Both BSSs and BLs are key to understanding the long-term effects of binary evolution, angular momentum transfer, and mass exchange in stellar populations. While most known examples reside in star clusters, Galactic field counterparts have also been identified \citep{Bond1971ApJ...165...51B, Bhat2025A&A...700L..23B}, suggesting that similar processes operate in diverse Galactic environments. Comprehensive reviews by \citet{Mathieu2025ARA&A..63..467M} and \citet{Wang2026enap....2..449W} summarize the growing importance of these systems as probes of stellar interaction and rejuvenation.

The first dedicated search for BLs was conducted by \citet{Leiner2019ApJ...881...47L}, who used K2 light curves and $v \sin i$ measurements to identify rapidly rotating main-sequence stars in M\,67. \citet{Jadhav2019ApJ...886...13J} subsequently identified another BL in the same cluster based on the presence of an extremely low-mass white dwarf companion, which is a clear signature of past mass transfer.
Since then, several clusters have been shown to host BL candidates \citep{Subramaniam2020JApA...41...45S, Dattatrey2023MNRAS.523L..58D, Narayan2025arXiv250912315S}. However, no large-scale survey exists due to the rarity of these objects and the limited number of well-studied host clusters. Chemical signatures of accreted material could also reveal such systems, though no systematic abundance study has yet been undertaken.

There are $>$4000 and $>$2000 BSSs known among Galactic open and globular clusters, respectively \citep{Simunovic2016MNRAS.462.3401S, Rain2021A&A...650A..67R, Jadhav2021MNRAS.507.1699J, Li2023A&A...672A..81L, Carrasco2025A&A...699A.142C}. Assuming BSSs and BLs are formed by similar mechanisms, their population sizes should be similar within a cluster.
This means that there should be thousands of BLs lurking along the main sequences.

In this work, we aim to present a formal definition of BLs and compile a list of BL candidates. Similar to \citet{Leiner2019ApJ...881...47L}, we
identify fast-rotating main-sequence stars that are potential BL candidates, using rotation catalogs derived from TESS, Kepler, and spectroscopic observations of open clusters.

\begin{figure*}
    \centering
    \includegraphics[width=0.98\linewidth]{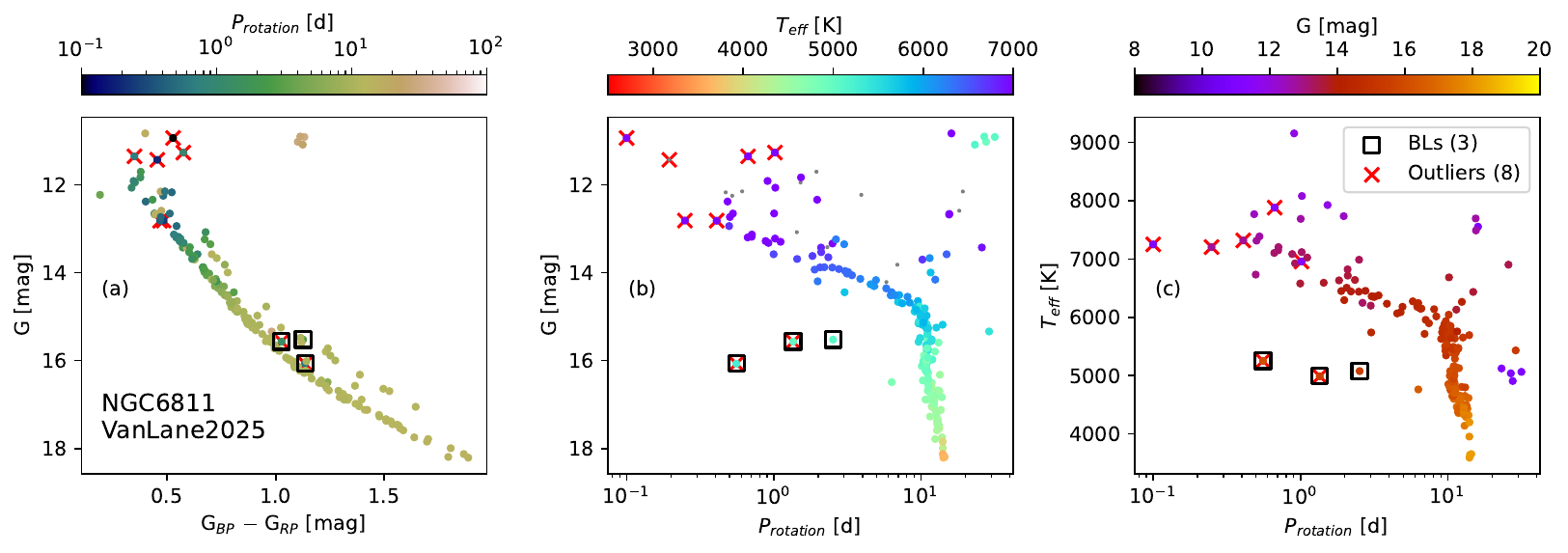}
    \caption{Diagnostic plots for NGC 6811.
    (a) The apparent \textit{Gaia} CMD colored according to the rotation period \citep{Van2025ApJ...986...59V}.
    (b) Distribution of $P_{rotation}$ and \textit{Gaia} $G$~magnitude colored according to the temperature (if unavailable, the stars are shown as gray dots).
    (c) Distribution of $P_{rotation}$ and temperature colored according to the G magnitude.
    The BL candidates (black squares) and the automatically detected outliers (red crosses) are highlighted.}
    \label{fig:diagnostic_combo}
\end{figure*}

\section{The definition} \label{sec:definition}

BSSs are defined observationally as stars bluer and brighter than the cluster turnoff. However, the terms `bluer', `brighter', and even `turnoff' lack strict quantitative definitions. Moreover, the choice of CMD can influence which stars are classified as BSSs; for instance, UV-based CMDs used in globular clusters yield different selections than optical CMDs \citep{Raso2017ApJ...839...64R}. BSSs can have various ages (100 Myr to 13 Gyr) and masses (10 to 0.9 M$_{\odot}$) depending on the properties (age and MSTO mass) of their host clusters \citep{Jadhav2021MNRAS.507.1699J, Carrasco2025A&A...699A.142C} and they can be single or part of binary or higher-order systems. 

Furthermore, the term `field-BSS' is used in the literature. These stars are typically identified through chemical peculiarities and age diagnostics, such as halo-like kinematics or low metallicity, that are inconsistent with their CMD positions \citep{Bond1971ApJ...165...51B, Panthi2023MNRAS.525.1311P}. In such cases, the BSS moniker denotes evidence of a past mass-accretion event, even though no reference population (like a cluster) is available to define their relative bluer or brighter nature.

Compared to BSSs, BLs are a more recently proposed class of objects, loosely defined and typically characterized by their method of detection rather than by strict physical criteria.
A BL can be identified as
i) a fast rotator on the main sequence \citep{Leiner2019ApJ...881...47L}; 
ii) a main-sequence star hosting companions that can only form through mass transfer (specifically via mass accretion by the BL component; \citealt{Jadhav2024A&A...688A.152J}); 
iii) a main sequence star showing chemical signatures of past accretion, such as Li/C/O depletion or s-process elements enhancement. 
However, an individual BL candidate may not satisfy all of these criteria, similar to how not all BSSs meet every diagnostic test.

To further complicate things, \citet{Bhat2025A&A...700L..23B} recently proposed a term `field-pre-BL' (a BL progenitor in the field) for a low-mass system expected to undergo future mass transfer. However, stellar mass alone cannot be used to differentiate between a field-BSS and a field-BL without a clear and unified definition. 
Similar to the BSSs, BLs also have a range in mass (smaller than the cluster turnoffs) and age (all possible cluster ages from a few Myr to several Gyr).
In fact, a BL in a young cluster can be more massive than a BSS in an old cluster. Furthermore, despite their name, BLs are not necessarily bluer in color, highlighting the need for a more physically motivated definition.

To decouple the definitions of BSSs and BLs from dependencies on mass, color, or host cluster properties, we propose the following physically motivated definitions\footnote{Similarly, a `yellow straggler' would be a post-main sequence subgiant/giant star that has gained mass through mergers, collisions, or mass transfer among relatively coeval stellar components.}:
\begin{itemize}
    \item Blue straggler:
    A H-core-burning star that has gained mass through mergers, collisions, or mass transfer among relatively coeval stellar components, such that a single star of comparable mass \textit{would have already evolved off the main sequence} according to the standard single stellar evolutionary theory.
    \item Blue Lurker:
    A H-core-burning star that has gained mass through mergers, collisions, or mass transfer among relatively coeval stellar components, such that a single star of comparable mass \textit{would still be on the main sequence} according to the standard single stellar evolutionary theory.
\end{itemize}

Only the BSSs and BLs lying within clusters should be referred to as BSSs and BLs, while those in the field should be referred to as field-BSSs and field-BLs.
These definitions are not perfect, as confirming H-core burning or accurately estimating the masses of post-interaction stars remains non-trivial, and no single stellar evolution model is without limitations. 
And the unrestricted definitions encompass all stellar interactions resulting in a H-core burning star, classifying them as either a BSS or a BL.
Nevertheless, the proposed definitions provide a useful conceptual framework for distinguishing between BSSs, field-BSSs, BLs, and field-BLs. 

Differentiating between field-BLs and field-BSSs further requires knowledge of the progenitor ages.
The most easily identifiable field-BLs are likely to be those that can be dynamically associated with a star cluster, such as ejected members or stars belonging to tidal tails, where the cluster age provides an independent constraint on the system’s age.

Note that these definitions assume that all stellar components were coeval, as is expected for the majority of stellar interactions. However, in rare cases, dynamical captures can produce binaries composed of stars with different ages \citep{Gomez2012ApJ...745...58G, Malkov2022OAst...31..327M}. Such systems may later interact, undergo rejuvenation, and form analogues of BSSs/BLs. We leave the naming of such systems to their discoverers.

\section{Blue lurker selection} 

\begin{table*}[]
    \caption{The catalog of new (97) and previously known (36) BL candidates.}
    \centering
\begin{tabular}{rllllrrc}
\toprule
\textit{Gaia} DR3 source\_id & Cluster & Rotation Ref. & Detection Ref. & comment & $v\sin i$ & Prot & ...\\
 &  &  &  &  & [km s$^{-1}$] & [d] & ... \\
\midrule
2320702191604400384 & Blanco\_1 & Healy2023 & This work &  &  & 0.3077$\pm$0.0083 & ...\\
249208987660681216 & Melotte\_20 & VanLane2025 & This work &  &  & 0.361& ... \\
2893942233835073792 & NGC\_2243 & GES & This work & RV\_outlier & 115$\pm$30 &  & ...\\ \hline
604917663714774784 & NGC\_2682 &  & Leiner2019 &  &  & 5.6& ... \\
342918645704887040 & NGC\_752 &  & Jadhav2024 & UV\_excess &  &  & ...\\
... & ... & ... & ... & ... & ... & ... & ...\\

\bottomrule
\end{tabular}
    \tablefoot{Full table with additional columns (e.g., \textit{Gaia} astrometry, photometry, best SED-fit parameters, and Simbad information) is available at CDS. Sources of rotational information: \citet{Geller2010AJ....139.1383G, Douglas2019ApJ...879..100D, Curtis2020ApJ...904..140C,  Fritzewski2023A&A...674A.152F,Healy2023ApJ...944...39H, Linck2024AJ....168..205L,Sha2024ApJ...977..103S, Van2025ApJ...986...59V}; GES \citep{Bragaglia2022A&A...659A.200B,Hourihane2023A&A...676A.129H}.
    Sources of membership information: \citet{Cantat2020A&A...640A...1C,Rao2023MNRAS.526.1057R,Hunt2024A&A...686A..42H}.
    Sources of previous BL classification: \citet{Jadhav2019ApJ...886...13J, Leiner2019ApJ...881...47L, Subramaniam2020JApA...41...45S, Dattatrey2023MNRAS.523L..58D, Jadhav2023A&A...676A..47J, Jadhav2024A&A...688A.152J, Panthi2024MNRAS.52710335P, Narayan2025arXiv250912315S}.
    }
    \label{tab:catalog}
\end{table*}

\subsection{Identification based on stellar rotation}

We compiled stellar rotation ($P_{rotation}$ and $v\sin i$) from multiple literature sources that analyzed one or more open clusters. The sample includes data from WIYN spectroscopy \citep{Geller2010AJ....139.1383G, Linck2024AJ....168..205L}, TESS light curves \citep{Healy2023ApJ...944...39H, Sha2024ApJ...977..103S}, Kepler/K2 light curves \citep{Douglas2019ApJ...879..100D, Curtis2020ApJ...904..140C}, and STELLA telescope light curves \citep{Fritzewski2023A&A...674A.152F}.
We further incorporated the literature-based rotation catalog of \citet{Van2025ApJ...986...59V}, which compiled period measurements derived from light-curve analyses using TESS, Kepler, K2, and Blanco telescope observations.
Additionally, we crossmatched the \citet{Cantat2020A&A...640A...1C} and \citet{Hunt2024A&A...686A..42H} open cluster membership catalogs with \textit{Gaia}-ESO DR5.1 \citep{Bragaglia2022A&A...659A.200B, Hourihane2023A&A...676A.129H} to facilitate $v\sin i$ based selection.
The catalogs were culled as recommended to remove invalid data or non-members.

We created diagnostic figures similar to Fig~\ref{fig:diagnostic_combo} for all clusters. 
Figs~\ref{fig:Prot_vs_G} and \ref{fig:vsini_vs_G} show the primary diagnostic plot for all clusters where BL candidates are identified via rotation. Note that the position of BLs known in the literature (which can be slow-rotating) is also highlighted in Figs~\ref{fig:Prot_vs_G} and \ref{fig:vsini_vs_G} (e.g., in NGC 2682).

The BL candidates were selected based on their fast rotation among cluster members with either $P_{rotation}<5$~days or $v\sin i>30$~km~s$^{-1}$.
We also used an automated outlier detection algorithm (\textsc{IsolationForest} from \textsc{sklearn}) to aid the outlier identification. However, no outlier detection algorithm is perfect, and they are heavily affected by the choice of various hyperparameters. We visually inspected the diagnostic plots with these outliers and manually created a list of BL candidates by adding or removing the automatically detected outliers. Radial velocity outliers ($>3$ standard deviations from the cluster mean value) were noted but not removed (discussed more in Section~\ref{sec:discussion}).

The detailed list of BL candidates is given in Table~\ref{tab:catalog}.
We also list the previously identified BL candidates reported in the literature (bottom part of Table~\ref{tab:catalog}). 
Of the total 133 BL candidates, 12 are detected in X-ray, 35 have excess UV flux (18 known from literature and 28 identified in Section~\ref{sec:seds}), and 21 are eclipsing and spectroscopic binaries. 
Six outliers were classified as young stellar objects in the literature and were removed from the BL candidates. In addition, \citet{Bhat2025A&A...700L..23B} mentioned a field pre-BL candidate (\textit{Gaia}~DR3~4265540383431508736) which is not part of the list. 

\begin{figure}
    \centering
    \includegraphics[width=0.98\linewidth]{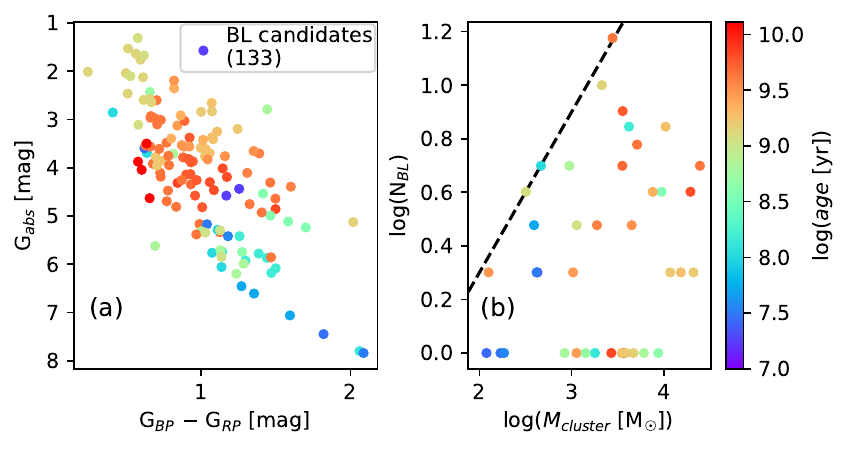}
    \caption{(a) Absolute \textit{Gaia} CMD of the 133 BL candidates.
    (b) Distribution of the number of BLs and the cluster mass. The points are colored according to cluster age. The dashed line showing the upper limit on the number of BSSs for given cluster mass is taken from \citet[eq. 1]{Jadhav2021MNRAS.507.1699J}. The 4 BLs in the globular cluster NGC 362 were assigned an age of 13 Gyr in panel (a) for visual convenience.}
    \label{fig:cmd}
\end{figure}

\subsection{Spectral energy distributions} \label{sec:seds}

To create spectral energy distributions (SEDs) of the BL candidates, we used pre-computed crossmatches \citep[which propagated proper motions, e.g.,][]{Marrese2017A&A...607A.105M, Bianchi2020ApJS..250...36B} to compile photometry from 
2MASS \citep{Cutri2003yCat.2246....0C}, 
GALEX GUVcat\_AIS GR6+7 \citep{Bianchi2017ApJS..230...24B}, 
PanSTARRS DR1 \citep{Chambers2016arXiv161205560C}, 
SDSS DR16 \citep{Ahumada2020ApJS..249....3A}, 
and 
WISE (\citealt{Cutri2014yCat.2328....0C}). 
Additionally, we did $\leq$1\arcsec\ on-sky crossmatch with
UVIT/\textit{AstroSat} \citep{Jadhav2021MNRAS.503..236J,Jadhav2023A&A...676A..47J, Jadhav2024A&A...688A.152J, Piridi2024ApJS..275...34P}
and
UVOT/\textit{Swift} \citep{Siegel2019AJ....158...35S} catalogs.
Most sources have optical and IR data, while only 40 have detection in at least 1 UV filter. Synthetic photometry for \textit{UVBRIugrizy}, F606W, and F814W passbands from \textit{Gaia} DR3 \citep{Montegriffo2023A&A...674A..33G} was used to augment any available data.
A few SEDs were not fitted due to either insufficient data or due to spatial crowding, which can lead to photometric contamination in non-\textit{Gaia} filters.

The optical-IR SEDs were fitted with Kurucz--UVBLUE hybrid model \citep{Castelli2003IAUS..210P.A20C, Rodriguez2005ApJ...626..411R}. The hybrid spectral models use the UVBLUE model for wavelengths $<4700$ \AA, and the Kurucz model in the redder region. The optical--IR SEDs were fitted using \textsc{Binary\_SED\_Fitting}\footnote{\url{https://github.com/jikrant3/Binary_SED_Fitting}}  \citep{Jadhav2021JApA...42...89J,jadhav_2024_13928317} assuming that the stellar distances, extinctions, and metallicities are the same as the host cluster. The UV points were deliberately ignored in the fitting process to compare the observed UV flux with the expected flux of a single star.

The BL candidates span 4000--9000 K in $T_{eff}$ and 0.5--3.2 R$_{\odot}$ in radii.
The best-fitted SED parameters for sources with enough data and reliable fits are listed in Table~\ref{tab:catalog}.
We detected significant fractional residual [$(F_{obs}-F_{model})/F_{obs}\geq0.3$] in at least one UV filter for 28 BL candidates (note that UV excess was already noticed in the literature for many of these sources). Fig~\ref{fig:SEDs_excess} shows the SEDs of the 6 BL candidates with excess flux in at least 3 UV filters. The number of UV excess filters is given in Table~\ref{tab:catalog}.

\section{Discussion} \label{sec:discussion}

Fig~\ref{fig:cmd} (a) shows the absolute \textit{Gaia} CMD of the BL candidates. Their (photometric) masses are in the range of 0.5--3 \Msun. The host open clusters have an age range of 18 Myr to 7 Gyr \citep{Cantat2020A&A...640A...1C} and a mass range of 120 to 25000 \Msun\  \citep{Hunt2024A&A...686A..42H}. 
Fig~\ref{fig:cmd} (b) shows the number of BLs and the cluster mass. The dashed line is the empirical upper limit on the number of BSSs in a cluster \citep[eq. 1]{Jadhav2021MNRAS.507.1699J}. The agreement with the BSS limit for the BLs shows that the number of BSSs and BLs has a similar dependence on the cluster mass.

Confirming genuine BLs requires the detection by more than one method of identification. 
Cross-verification between photometric and spectroscopic rotation indicators, using light-curve analysis of $v\sin i$ outliers and $v\sin i$ measurements of photometric outliers, will help confirm the rapid rotation.
As mentioned in \S~\ref{sec:definition}, BLs can also be identified through diagnostics unrelated to rotation. Characterizing their companions, such as extremely low-mass white dwarfs, via spectroscopy or spectral energy distributions provides a powerful alternative approach.
Similarly, chemical peculiarities such as s-process element enrichment or Li/C/O depletion in main-sequence stars may indicate a past episode of mass transfer. 
Similar to BSSs, most BLs are expected to reside in binary systems; therefore, radial velocity monitoring to identify spectroscopic binaries and determine their orbital parameters and companion properties is essential. Ultraviolet observations will further aid in detecting and characterizing hot compact companions, providing direct evidence of past mass transfer. The 28 BL candidates with UV excess are perfect candidates for such analysis.

A subset of the BL candidates shows radial velocities inconsistent with their host clusters (highlighted in Table~\ref{tab:catalog}). These stars could be non-members or part of binary systems, which appear as outliers due to the orbital motion. Further RV monitoring is required to confirm membership and, for spectroscopic binaries, to determine their orbital and stellar parameters.

\subsection{Contamination}

Apart from mass accretion, stars can be fast-rotating due to other reasons and thus could represent false positives in our sample.
i) Most intermediate stellar mass stars ($\approx 1.3$--8~M$_\odot$) are born fast rotators \citep{Cordoni2024MNRAS.532.1547C}, and they do not slow down significantly during their main sequence lifetimes due to radiative envelopes, unless they are in short-period binary systems. Therefore, their maximum $v \sin i$ is observed to range from 100~km~s$^{-1}$ to 400~km~s$^{-1}$.
ii) Pre-main-sequence stars are also expected to spin up when they contract on the main sequence \citep{Bouvier2014prpl.conf..433B}.
The candidate list contains 10 BLs in clusters younger than 100 Myr. These objects may be intrinsic fast rotators due to their youth rather than past mass-transfer events.
iii) High stellar rotation can arise from tidally locked close binaries without any history of mass transfer \citep{Moraux2013A&A...560A..13M}. 
iv) Rotation measurements can be affected by contamination from solar or lunar cycles \citep{Getman2023ApJ...952...63G}, or by light-curve misclassification.
v) Minimal magnetic breaking in stars hotter than the Kraft break at 6200 K \citep{Schatzman1962AnAp...25...18S, Kraft1967ApJ...150..551K, Beyer2024ApJ...973...28B}.
vi) Erroneous rotation ($P_{rotation}$ or $v\sin i$) measurements.
vii) Misclassification of a younger star as a cluster member.
Improved membership catalogs and spectroscopic follow-up will be essential for refining the sample and quantifying its purity.

\subsection{Completeness}

\subsubsection{Missing BLs due to the rotation detectability}

The rotation-based search is sensitive only to stars exhibiting detectable photometric modulation or spectroscopic rotational broadening. Consequently, the present list excludes stars with minimal surface spot coverage, resulting in undetectable light-curve periodicity, or with low inclination angles leading to small $v\sin i$. Blue stragglers are known to spin up after mass transfer and subsequently spin down with time \citep{Nine2023ApJ...944..145N}. 
A similar scenario likely applies to BLs, which may lose angular momentum and slow their rotation.
The severity of the slowdown will be determined by the mass of the BL: the lower mass BLs with convective envelopes will slow down faster compared to more massive BLs with radiative envelopes.
Determining the cooling ages of white dwarf companions to BLs, in various mass ranges, offers a direct way to test this hypothesis.
If there are too many rapidly rotating BLs in a cluster, then they would not appear as outliers and would be missed in this work. More theoretical effort is required to define the nominal rotation periods of stars across evolutionary phases (e.g., pre-main sequence, split main sequence, or extended MSTO) and to establish thresholds for identifying anomalously rapid rotators.

\subsubsection{Missing BLs due to selection effects}

Our identification process depends on cluster membership determination, which itself may include false positives and negatives. Moreover, since BLs are likely to reside in binaries as suggested by the high binary fraction among BSSs, their astrometric solutions in \textit{Gaia} can show excess noise, leading to their exclusion from membership catalogs (see discussion in \citealt{Tagaev2025arXiv250910435T} regarding the \citealt{Hunt2024A&A...686A..42H} catalog).

\subsubsection{Completeness of the $v\sin i$ based selection}

Using \textit{Gaia} DR3–RVS and \textit{Gaia}-ESO DR5.1 data, 199 out of 274 BSSs with measured $v\sin i$ values are rapid rotators ($v\sin i > 30$ km s$^{-1}$). Assuming BLs exhibit similar rotational properties, an ideal rotation-based census would detect roughly 72\% of them.
However, as not all high-$v\sin i$ stars are BLs, and our list excludes many fast rotators from intermediate-mass stars (1.5 -- 8~\Msun) to avoid contamination, the effective completeness of the presented BL catalog is $<72\%$. 
Conservatively assuming that 20\% of the member stars are missing in \textit{Gaia} membership catalogs \citep{Tagaev2025arXiv250910435T}, the completeness reduces to $<58\%=72\times0.8\%$.
Furthermore, only about $\approx$5\% of cluster main-sequence stars currently have $v\sin i$ measurements (based on \citealt{Hunt2024A&A...686A..42H} and \citealt{Cantat2020A&A...640A...1C} membership and \textit{Gaia}-ESO, LAMOST, APOGEE, and GALAH surveys). Thus, the overall completeness is $<2.9\%=72\times0.8\times0.05\%$.
The roughly 60 $v\sin i$-based BL candidates identified in this work suggest a total BL population of approximately $>$2000, comparable to the current census of known BSSs.
Since BLs remain on the main sequence for longer timescales after formation and can occupy the full extent of the main sequence, their intrinsic population could in fact exceed that of BSSs.
This highlights the fact that rotation-based census is powerful in identifying BL candidates; however, it is primarily limited by the data availability and quality.

\section{Conclusions and summary}

This study conducted an observational survey of BL candidates in open clusters. Overall, we identified 97 new BL candidates across 35 open clusters using rotation measurements derived from TESS, Kepler, K2, and \textit{Gaia}-ESO survey data. Including 36 previously known BL candidates from five open clusters and one globular cluster, the total number of known or suspected BLs now stands at 133.

Based on clusters with significant BL and BSS populations, we expect their numbers to be comparable. Given that more than 4000 open cluster BSSs are currently known and the estimated $\approx$3\% completeness of the present BL catalog, thousands of BLs likely remain undetected within Galactic clusters. Together, BLs and BSSs may constitute 5--10\% of the total cluster population, underscoring their importance in refining our understanding of single and binary stellar evolution.

Expanding TESS-based rotation measurements for clusters and wide-scale spectroscopic data will be crucial to identifying additional BLs, and similar analyses of cluster tidal tails could reveal field BLs dynamically linked to their parent clusters. Known binary BL candidates warrant targeted follow-up, while coordinated time-series, spectroscopic, and UV observations will be essential to confirm their mass-transfer histories and establish a comprehensive census of these post-interaction systems.

As most BSS progenitors are stars near the MSTO, their main-sequence lifetimes are typically extended as a result of mass accretion \citep{Wang2026enap....2..449W}.
In contrast, BLs may gain sufficient mass to accelerate their H-burning, thereby shortening their main-sequence phase. Hydrodynamical simulations exploring a grid of accretor birth masses, accreted mass, the H-content of the accreted material, the internal mixing, and epochs of mass transfer will be instrumental in identifying the regions of parameter space that lead either to lifetime extension following accretion or to a more rapid evolutionary demise.

Overall, understanding the frequency and evolution of such interacting binaries is key to refining models of stellar populations. These systems highlight the limitations of single-star evolutionary assumptions and emphasize the need for binary-inclusive frameworks to explain the observed properties of resolved and unresolved populations.

\section*{Data Availability}

The full version of Table~\ref{tab:catalog} is available in electronic form at the CDS via anonymous ftp to \url{https://cdsarc.u-strasbg.fr/} (130.79.128.5) or via \url{http://cdsweb.u-strasbg.fr/cgi-bin/qcat?J/A+A/}.

\begin{acknowledgements}
    We thank the anonymous referee for quick and constructive comments. 
    ER acknowledges the support from the European Union (ERC-2022-AdG, {\em "StarDance: the non-canonical evolution of stars in clusters"}, Grant Agreement 101093572, PI: E. Pancino). Views and opinions expressed are, however, those of the author(s) only and do not necessarily reflect those of the European Union or the European Research Council. Neither the European Union nor the granting authority can be held responsible for them.
    The work used the following tools for the analysis: 
    \textsc{Astropy} \citep{astropy:2022}; 
    \textsc{Astroquery} \citep{Ginsburg2019AJ....157...98G};
    \textsc{Matplotlib} \citep{Hunter:2007};
    \textsc{NumPy} \citep{2020Natur.585..357Harris}; 
    \textsc{Scikit-learn} \citep{Pedregosa2011JMLR...12.2825P};
    \textsc{topcat} \citep{2005ASPC..347...29TOPCAT}.
\end{acknowledgements}

\bibliographystyle{aa}
\bibliography{references}

\begin{appendix}

\section{Supplementary table and figures}

\begin{figure*}
    \centering
    \includegraphics[width=0.98\linewidth]{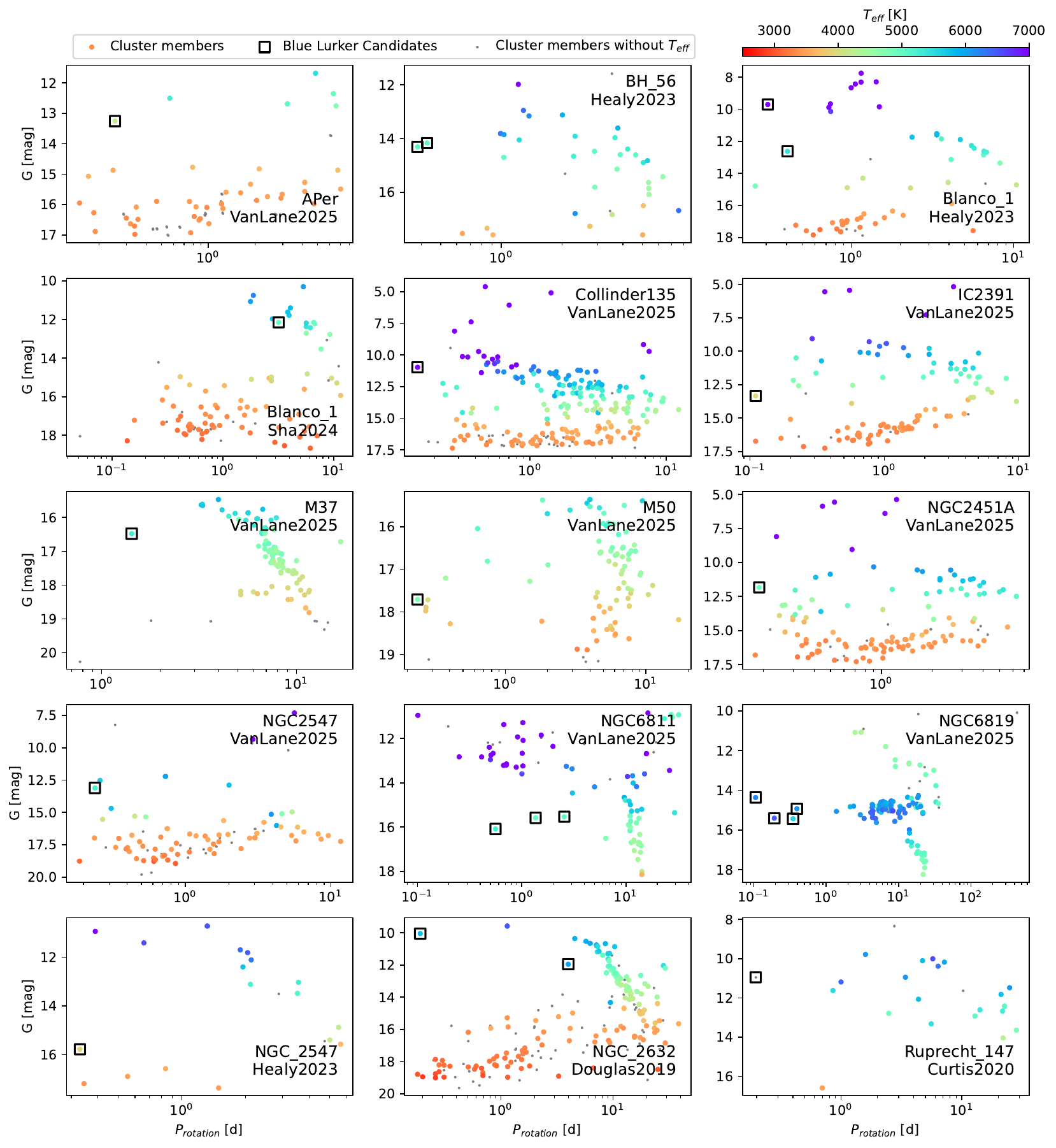}
    \caption{Distribution of the rotation periods and \textit{Gaia} $G$~magnitudes for clusters analyzed with light curves.
    The stars are colored by their temperature (if unavailable, the stars are shown as gray dots). The BL candidates are highlighted by the black squares.}
    \label{fig:Prot_vs_G}
\end{figure*}

\begin{figure*}[!h]
    \centering
    \includegraphics[width=1\linewidth]{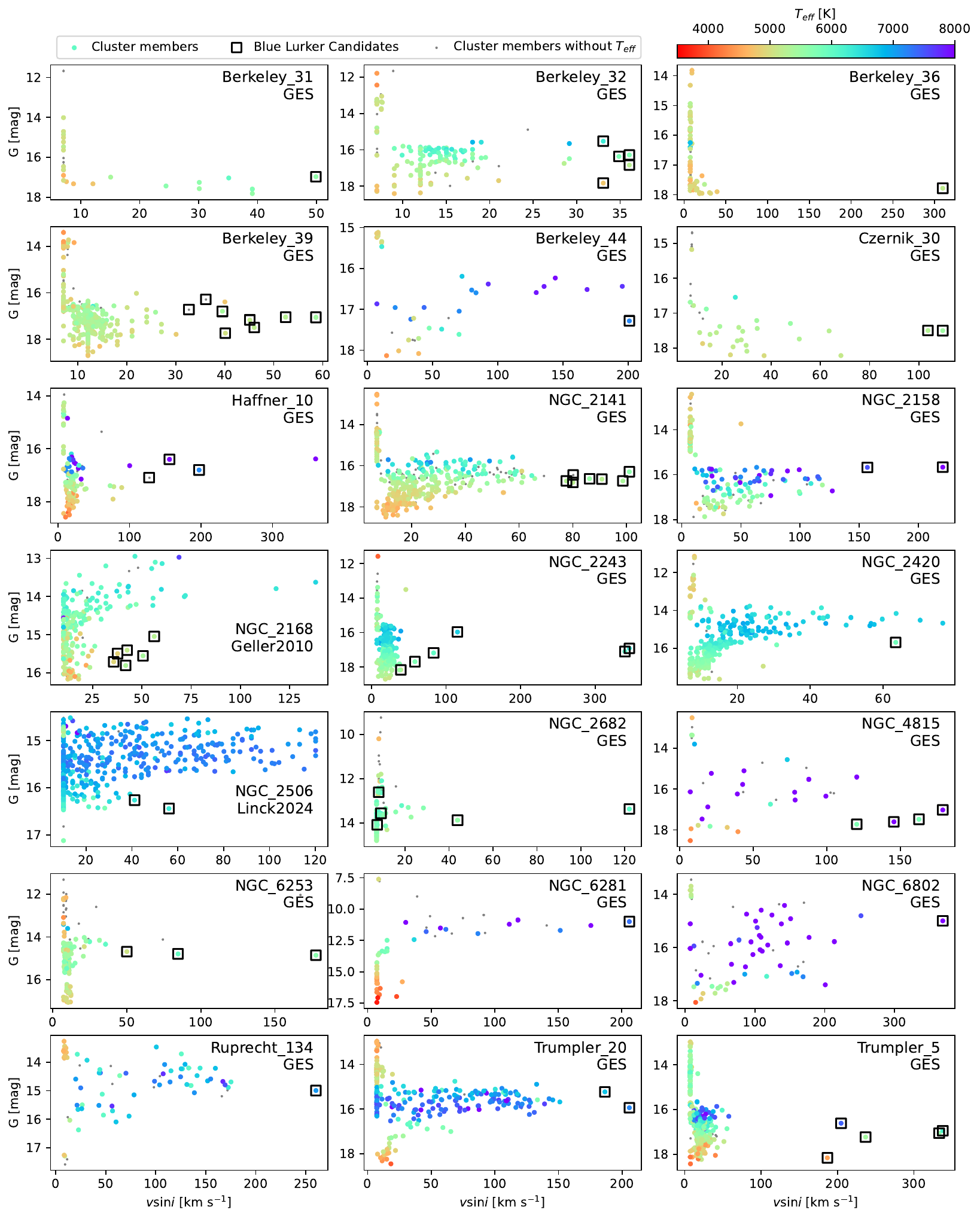}
    \caption{Distribution of the $v\sin i$ and \textit{Gaia} $G$~magnitudes for clusters analyzed with spectroscopy.
    The stars are colored by their temperature (if unavailable, the stars are shown as gray dots). The BL candidates are highlighted by the black squares.}
    \label{fig:vsini_vs_G}
\end{figure*}

\begin{figure*}
    \centering
    \includegraphics[width=0.98\linewidth]{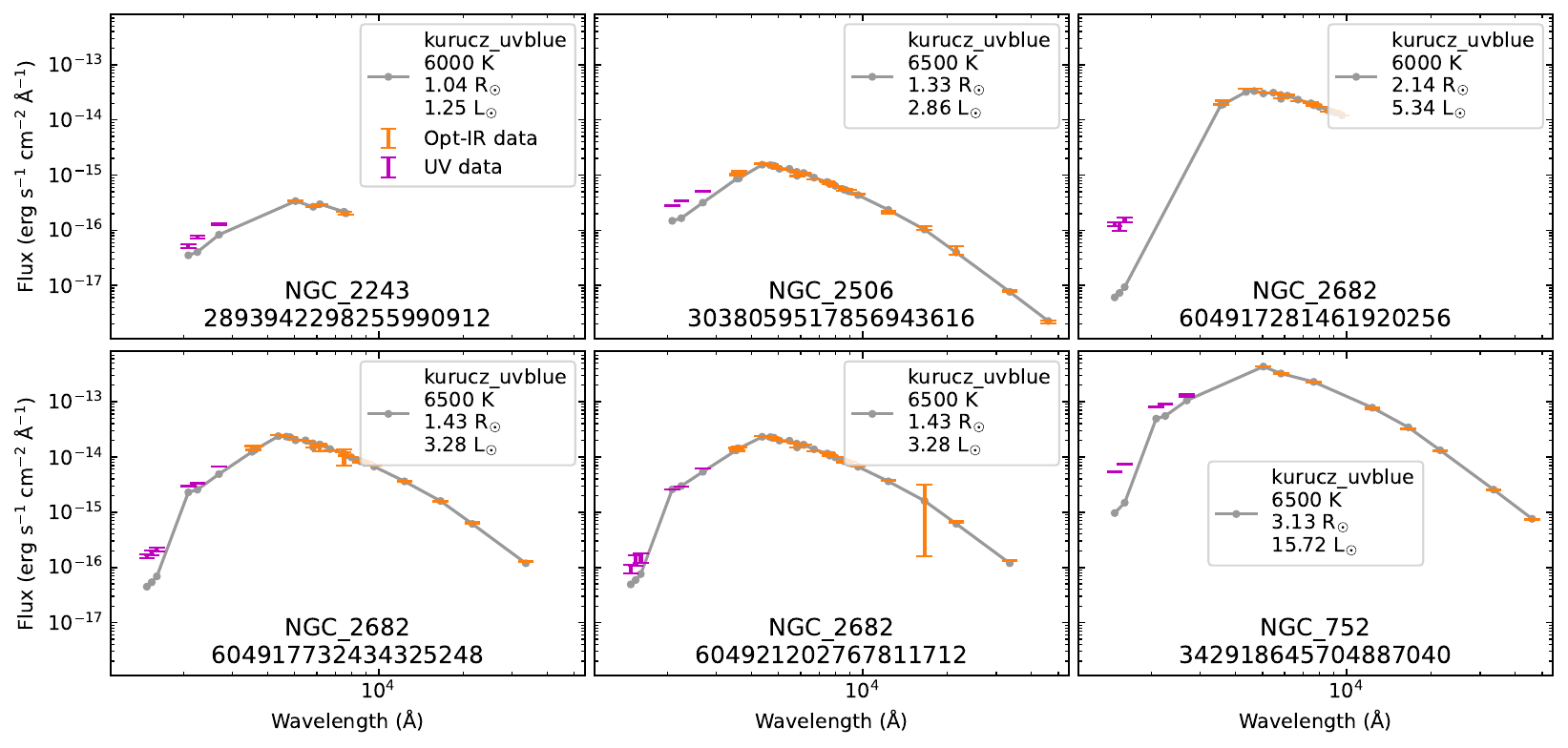}
    \caption{SEDs of stars with significant excess flux in at least 3 UV filters. The UV (magenta) and optical--IR fluxes (orange) are shown using the corresponding flux errors. The best-fitted Kurucz--UVBLUE model is shown in gray.}
    \label{fig:SEDs_excess}
\end{figure*}

\end{appendix}

\end{document}